\documentstyle[12pt]{article}
\begin{document}
\baselineskip 18pt
\newcommand{\Dirac}{/\!\!\!\!D}
\newcommand{\beq}{\begin{equation}}
\newcommand{\eeq}[1]{\label{#1}\end{equation}}
\newcommand{\bea}{\begin{eqnarray}}
\newcommand{\eea}[1]{\label{#1}\end{eqnarray}}
\renewcommand{\Re}{\mbox{Re}\,}
\renewcommand{\Im}{\mbox{Im}\,}
\newcommand{\La}{\Lambda}
\newcommand{\tr}{\mbox{tr}\,}
\def\cP{{\cal P}}
\def\cF{{\cal F}}

\begin{titlepage}
\hfill  hep-th/9704163 NYU-TH-97/03/02 IASSNS-HEP-97/25 IFUM/FT-558
\begin{center}
\hfill
\vskip .4in
{\large\bf Spontaneously Broken N=2 Supergravity Without Light Mirror Fermions}
\end{center}
\vskip .4in
\begin{center}
{\large L. Girardello}
\vskip .1in
{\em Dipartimento di Fisica, Universit\`a di Milano\\ Via Celoria 16, 20133,
Milano, Italy\footnotemark}
\footnotetext{email girardello@milano.infn.it}
\vskip .1in
{\large M. Porrati}
\vskip .1in
{\em Department of Physics, NYU, 4 Washington Pl.,
New York, NY 10003, USA\\ and\\ Rockefeller University, New York, NY
10021-6399, USA\footnotemark}
\footnotetext{e-mail massimo.porrati@nyu.edu}
\vskip .1in
{\large A. Zaffaroni}
\vskip .1in
{\em Institute for Advanced Study, Olden Lane, Princeton, NJ 08540,
USA\footnotemark}
\footnotetext{e-mail zaffaron@sns.ias.edu}
\end{center}
\vskip .4in
\begin{center} {\bf ABSTRACT} \end{center}
\begin{quotation}
\noindent
We present a {\em spontaneously} broken N=2 supergravity model that reduces, in
the flat limit $M_{Planck}\rightarrow \infty$, to a globally supersymmetric
N=2 system with explicit soft supersymmetry breaking terms.
These soft terms generate a mass $O(M_W)$ for mirror quarks and leptons,
while leaving the physical fermions light, thereby overcoming one of the major
obstacles towards the construction of a realistic N=2 model of elementary
interactions.
\end{quotation}
\vfill
\end{titlepage}
\eject
\noindent
\section*{Introduction}
In N=2 supersymmetric four-dimensional theories, all non-gravitational
interactions are gauge interactions; therefore, N=2 is powerful enough to
relate the Yukawa couplings to the gauge coupling(s), which are instead
unrelated by N=1 supersymmetry. Exact N=2 supersymmetry also allows for the
derivation of exact non-perturbative results on the dynamics of gauge
theories~\cite{SW}; some of these results survive the explicit breaking of
N=2 by soft terms~\cite{AG}.

In spite of these attractive features, N=2 theories suffer from a serious
problem that has hampered their use as realistic models of elementary
interactions: all particles appear in {\em real} representations of the gauge
group. Clearly, to recover the particle content of the standard model, in
which particles belong to chiral (aka complex) representations of
$SU(2)\times U(1)$, something has to happen.

Two mechanisms for generating chirality are known. The first requires a higher
dimensional theory, which is itself chiral in higher dimensions. By
compactifying to four dimensions, one finds that the mass term of fermions
is given by the Dirac operator on the compact space. The number of chiral
families (better, the mismatch between chiral and anti-chiral families) is then
given by an index theorem. By wisely choosing the compactification, this
index may be nonzero. A pioneering example of such compactification was given
in~\cite{SS}. The most successful example is the well known Calabi-Yau
compactification of the heterotic string~\cite{CHSW}.
We must emphasize that these examples are four-dimensional theories with an
{\em infinite} number of fields, all but a finite number of them have masses
of the order of the inverse size of the compact dimensions. In other words,
in these theories, N=2 supersymmetry is broken at the compactification scale,
where the very notion of a four-dimensional space-time breaks down.

The second mechanism gives a very different scenario. There, the world is N=2
supersymmetric well below the compactification scale, or even the GUT scale;
thus, the effective four-dimensional theory is N=2, with a {\em finite}
number of fields. In this case, one can still recover the particle content of
the standard model by giving Majorana masses to the ``mirror'' fermions
belonging to the wrong-chirality representation of $SU(2)\times U(1)$. Since
Majorana masses necessarily break the gauge group, one finds a
model-independent constraint on the mirror masses: they must be of the order
of the $SU(2)\times U(1)$ breaking:
\bea
M_{mirror}\sim 100 \, GeV.
\eea{1}
The existence of mirror fermions is still compatible with experiment, for
appropriate mixing angles with physical fermions~\cite{NR}.
This model-independent constraint, at least, makes N=2 models interesting,
since they make a definite prediction that can be verified by future
experiments.

An N=2 model which implements this scenario was proposed in~\cite{DA}. Two
difficulties face anyone attempting to generate tree-level
mirror-fermion masses in N=2 supersymmetry.

The first, solved in ref.~\cite{DA}, is that physical fermions,
belonging to hyper-multiplets~\cite{F}, can only get tree-level masses
by the VEV of a complex
scalar, supersymmetric partner of the gauge field, and belonging to the
adjoint representation of the gauge group. This adjoint field must play the
role of the standard-model Higgs field. This fact requires an extension of
both the standard-model gauge group and the Higgs sector.
The minimal such extension, given in~\cite{DA}, is as follows.
\begin{itemize}
\item The gauge group is extended to $SU(3)\times SU(4)\times U(1)$.
\item $SU(4)\times U(1)$ is broken to $SU(2)\times U(1)$ by two Higgs
hypermultiplets in the $(1,4,+1/2)\oplus (1,\bar{4},-1/2)$ and
$(1,4,-1/2)\oplus (1,\bar{4},+1/2)$ of the gauge group, respectively.
\item Quarks and leptons, together with their mirrors, belong to real
representations of the gauge group. In the notations of ref.~\cite{DA}:
\bea
X_L=\left(\begin{array}{c} L \\ \overline{L}' \end{array}\right) \sim
(1,4,-1/2),
& &
Y_L=\left(\begin{array}{c} L' \\ \overline{L} \end{array}\right) \sim
(1,\bar{4}, +1/2), \nonumber \\
X_Q=\left(\begin{array}{c} Q \\ \overline{Q}' \end{array}\right) \sim
(3,4,+1/6),  & &
Y_Q=\left(\begin{array}{c} Q' \\ \overline{Q} \end{array}\right) \sim
(\bar{3},\bar{4}, -1/6).
\eea{2}
Here $Q$, $L$ denote the physical quarks and leptons, while $Q'$, $L'$ denote
the mirrors. Notice that a given irreducible representation of the gauge group
contains both physical fermions and mirrors.
\end{itemize}

The second difficulty proved harder to solve. In order to achieve the
right pattern of symmetry breaking, the authors of~\cite{DA} introduce by hand
some soft terms, which preserve the good ultraviolet properties of the N=2
theories~\cite{GG,PW}, but that, on the other hand, {\em explicitly} break
the N=2 supersymmetry. Ref.~\cite{DA} suggests that these terms may come from a
{\em spontaneously} broken N=2 supergravity, much in the same way as the
corresponding terms
arise in N=1 supergravity~(see~\cite{N} and references therein).

The quest for such a spontaneously broken N=2 supergravity has been elusive
so far. Indeed, even though N=2 supergravity models spontaneously broken to
N=1 exist~\cite{CGP,FGP,FGP2}, none has been found, which generates
mirror-fermion masses.

Purpose of this paper is to exhibit such a supergravity model. This model has
the field content of ref~\cite{DA}, supplemented with the minimal hidden
sector necessary to {\em spontaneously} break N=2 supersymmetry with two
independent scales. In the flat limit where the Planck mass
$M_{P}\rightarrow \infty$,
while the mass of both gravitini is kept constant, the SUSY breaking
generates soft terms (tri-linear Yukawa couplings and masses), which allow one
to
recover the model of ref.~\cite{DA} in an appropriate phenomenologically
realistic range of parameters. This model removes the major (though not
unique) obstacle to the construction of realistic models where N=2
supersymmetry is spontaneously broken well below the Planck (or
compactification) scale.

This paper is organized as follows: in Section 1 we review the basic facts
about supergravity Lagrangians, using the geometric formulation
of~\cite{DAFF}; in Section 2 we review the construction of the softly broken
N=2 model with tree-level mirror splitting of ref.~\cite{DA}. Section 3
describes how to spontaneously break N=2 supergravity with two independent
scales. Section 4 is the heart of the paper. There, we show how to recover
the soft supersymmetry breaking terms needed to give mass to the mirror
fermions from the flat limit of an N=2 supergravity. Some concluding remarks
are given in Section 5, while Appendix A contains an explicit construction of
the hypermultiplet manifold needed in Section 4.

The reader interested in the results, rather than in details of the
construction can skip Sections 1 and 3.

\section{N=2 Supergravity Lagrangians}
The fields of N=2 supergravity belong to the graviton multiplet, the vector
multiplet and the hypermultiplet. The graviton multiplet contains the graviton,
two gravitinos and a vector field. The vector multiplet contains a vector
field, a complex scalar and a Majorana fermion. The hypermultiplet contains
four real scalars and a Dirac fermion.

The bosonic part of the N=2 vector multiplet contains a complex scalar $z^i$
in addition to the gauge field; supersymmetry constrains the scalar to
parametrise a special K\"ahler
manifold of real dimension $2n_V$ (where $n_V$ is
the number of vector multiplets in the theory).
The special geometry of the vector-multiplet manifold is specified by
$2(n_V+1)$ holomorphic functions $X^\Sigma(z^i), F_\Sigma(z^i)$ ($i=1,..,n_V,
\Sigma = 0,..,n_V$)~\cite{df} , in terms of which the K\"ahler potential reads:
\beq
K = -\log i(X^{*\Lambda}F_{\Lambda} - X^{\Lambda}F^*_\Lambda).
\eeq{3}
The metric on the scalar manifold is $g_{ij^*} =
\partial_i\partial_{j^*}K$. In N=2 supergravity there is an extra vector field
which belongs to the graviton multiplet, called graviphoton;
no scalar fields is associated to it. Roughly speaking, the index $\La$ labels
 all vector fields ($n_V+1$), while the index $i$ labels the complex scalars
($n_V$) or, equivalently, the vector multiplets.

In a coordinate-independent definition of a special K\"ahler manifold, the
functions ($X^\La ,F_\La$) are holomorphic sections of a $2n_V+2$
dimensional
symplectic bundle over the manifold. The notion of a prepotential
$F(X)$ which characterizes the N=2 literature in the context of
supergravity~\cite{VP,df} as
well as in the rigid limit~\cite{SW}, can be recover only in particular
cases and it is associated with the choice of ``special'' coordinates.
The prepotential exists provided
that the matrix
\beq
e^a_i(z)=\partial_i\left ( {X^a\over X^0}\right)
\eeq{4}
is invertible. In this case $X^\Lambda$
can be regarded as a set of homogeneous
coordinates for the special K\"ahler manifold, and
\beq
F_\Lambda (X) = {\partial\over \partial X^\Lambda}F(X),
\eeq{5}
where the $F(X)$ is a homogeneous function of degree 2. Under these
circumstances one can use {\it special coordinates} $t^a=X^a/X^0$,
and the whole geometry
is encoded in a single holomorphic prepotential $(X^0)^{-2}F(X)$.

As shown in Section 4, we find a pattern of supersymmetry
breaking which splits the mirrors by using a  choice of
sections for which the prepotential does not exist; such sections can be
obtained, for example, by applying an appropriate symplectic transformation
to sections derived from a prepotential.

N=2 hypermultiplets contain four real scalars. N=2 supergravity requires
such scalars to be the coordinates of a quaternionic manifold~\cite{BW}.

A quaternionic manifold is a $4n_H$-dimensional real manifold with three
complex structures $J^x$ that satisfy the quaternionic algebra
\beq
J^xJ^y = -\delta_{xy} + \epsilon^{xyz}J^z, \;\;\; x=1,2,3,
\eeq{6}
such that the metric $ds^2=h_{uv}dq^udq^v, u,v=1,..,4n_H$ is
hermitian with respect to them. The two-forms $K_{uv}^x=h_{uw}(J^x)^w_v$
are covariantly closed with respect to an $SU(2)$ connection $\omega^x$
\beq
\nabla K^x \equiv d K^x + \epsilon^{x y z} \omega^y \wedge
K^z = 0.
\eeq{7}
To complete the definition of a quaternionic manifold we must impose that the
curvature $\Omega^x$ of the connection $\omega^x$ is proportional to the form
$K^x$
\beq
 \Omega^x \, \equiv \, d \omega^x +
{1\over 2} \epsilon^{x y z} \omega^y \wedge \omega^z= \lambda K^x.
\eeq{8}
The proportionality coefficient between $K$ and $\Omega$ is arbitrary;
the choice $\lambda= -M_P^{-2}$ gives the correct normalization of the kinetic
terms in the supergravity Lagrangian. Notice that the definition of a
hyperk\"ahler
manifold differs only in that $\Omega^x$ vanishes, rather than being
proportional to $K^x$. This corresponds to taking the limit
$M_P\rightarrow \infty$ in eq.~(\ref{8}).

The gauge group is a subgroup of the isometries of the total scalar manifold
parametrised by $z^i$ and $q^u$. Since there are $n_V+1$ vectors,
the gauge group
has dimension $n_V+1$; its action on the scalars is given by
$n_V+1$ Killing vectors:
\bea
z^i & \rightarrow & z^i+\epsilon^\Lambda k^i_\La (z),\nonumber\\
q^u &\rightarrow & q^u+\epsilon^\La k^u_\La (q)\ .
\eea{9}
To write the N=2 Lagrangian, one must introduce a triplet of real
prepotential for the quaternionic manifold~\cite{VP,df}.
They are the supergravity
generalization of the triplet of D-terms
of rigid N=2 supersymmetric gauge theories, and reduce to them in the flat
limit. They are defined as the  three real functions $P^x_\Lambda$ which
satisfy
\beq
2k^u_\La \Omega^x_{uv}=-\nabla_v P^x_\La=-(\partial_v P^x_\La+\epsilon^{xyz}
 \omega^y_v P^z_\La).
\eeq{10}

Now we have all the ingredients to write the (bosonic part of) the N=2
supergravity Lagrangian~\cite{df}:
\bea
{\cal L}_{bos} &=& -{1\over 2} R +
g_{ij^*}\nabla^\mu z^i \nabla_\mu \bar z^{j^*}+
h_{uv}\nabla_\mu q^u \nabla^\mu q^v +\nonumber\\
&&{\rm i}\left(
\bar {\cal N}_{\Lambda\Sigma}{\cal F}^{-\Lambda}_{\mu\nu}{\cal F}^{-\Sigma
\mu\nu} -
{\cal N}_{\Lambda\Sigma} {\cal F}^{+\Lambda}_{\mu\nu}{\cal F}^{+ \Sigma
{\mu\nu}}\right ) - V(z,\bar z, q),
\eea{11}
where  $\cF^{\pm\La}_{\mu\nu}={1\over 2} (\cF^\La_{\mu\nu}\pm
{{\rm i}\over2}\epsilon^{\mu\nu\rho\sigma}\cF_{\rho\sigma}^\La )$.
If we define
$(f_i^\Lambda, h_{i\Lambda}) = (\partial_i + \partial_i
K)(X^\Lambda,F_\Lambda)$, the matrix ${\cal N}_{\Lambda\Sigma}$ is
determined by
\beq
F_\Lambda = {\cal N}_{\Lambda\Sigma} X^\Sigma , h_{i^* \Lambda} =
{\cal N}_{\Lambda\Sigma} f^\Sigma_{i^*}.
\eeq{13}
${\cal N}$ is the scalar-dependent gauge kinetic term. When a prepotential
exists, it reduces to the
familiar~\cite{VP,SW} expression
${\partial^2F\over \partial X^\La\partial X^\Sigma}$.
The potential is given by:
\beq
 V( z, {\bar z}, q)
= e^K\left [\left(g_{ij^*} k^i_\Lambda k^{j^*}_\Sigma+4 h_{uv}
k^u_\Lambda k^v_\Sigma \right ) \bar X^\Lambda X^\Sigma
+ \left (g^{ij^*} f^\Lambda_i f^\Sigma_{j^*}
-3\bar X^\Lambda X^\Sigma\right ){\cal P}^x_\Lambda
{\cal P}^x_\Sigma \right ].
\eeq{12}

Since we are interested in the flat limit $M_P\rightarrow\infty$, we must
restore
physical normalizations in the previous Lagrangian, in which all the fields are
dimensionless. The correct assignment is to restore the right
dimension for the Lagrangian by multiplying it by $M_P^4$ and
to keep dimensionless the fields in the graviton
multiplets and in the hidden sector of the theory, while restoring dimensions
in the generic field $x$ in the physical sector by writing it as $x/M_P$.
The flat limit is then obtained by sending $M_P\rightarrow\infty$ while
keeping $x$ finite. The hidden sector will trigger supersymmetry breaking and
interference terms between the hidden and physical sectors will give rise to
soft supersymmetry breaking terms in the flat limit. These terms are
exhaustively discussed in Section 4. For the time being,
let us focus on the physical sector and let us work out
the simplifications in the previous formulas due to the flat limit.

Let us denote with $z^i$, $b^u$, respectively, the scalar
partners of the gauge fields and the hypermultiplets scalars in the hidden
sector, and denote with $\phi^i,q^u$ the same quantities in the physical
sector. Since the kinetic term for the hidden-sector
scalars is proportional to $M_P^2$,
in the flat limit, all hidden-sector scalars ``freeze'' to their
vacuum expectation value.
Also, when $M_P\rightarrow\infty$, all the previous quantities can be expanded
in powers of $\phi^i,q^u$. A standard dimensional argument says that in the
physical sector  only renormalizable terms will survive in this limit.

By expanding the K\"ahler potential and the quaternionic metric in inverse
powers of $M_P$,
and keeping only the leading order, all
the scalar metrics become obviously flat. By appropriately choosing
the supergravity metric, all observable-sector fields will be canonically
normalized in the flat limit.

First, Let us describe the hypermultiplets.
The metric is flat and can be normalized
to $h_{uv} = \delta_{uv}$.
The quaternionic geometry reduces to
the hyperk\"ahler one in the flat limit, since $\Omega^x$ is proportional to
$1/M_P^2$.

It is convenient to represent the four scalars $q^u$ in a hypermultiplet as
a quaternion
\bea
 Q &=& e_u q^u = \left (\begin{array}{cc} x & -y^*\\ y & x^*\end{array}\right),
\nonumber\\
q^u &=& {1\over 2}\tr \bar e^u Q,
\eea{14}
where $e_u =(1,-i\vec\sigma ), \bar e^u =(1,i\vec\sigma )$.

There is an alternative representation of the hypermultiplets in the N=2
literature~\cite{VP} in which the scalar fields $A_i^a$ have an index $i$,
denoting that it is a doublet of the $SU(2)$ global R-symmetry, and
an extension index which transforms under the gauge group. They must satisfy
the reality constraint:
\beq
\left ( A_i^a\right )^* = \epsilon^{ij}\rho_{ab}A_j^b,
\eeq{15'}
where $\rho\rho^* = -1$ for consistence. This can be solved only if
the space labeled by $a$ is even-dimensional.
A convenient choice for  $\rho$
is a block diagonal form in which the entries are  $-i\sigma_2$. For a single
hypermultiplet the solution of the reality constraint is exactly the
quaternion in eq.~(\ref{14}).

The linear action of the gauge group,
\beq
\delta^\La Q = -i{\cal T}^\La Q,
\eeq{15}
is constrained by the reality condition to:
\beq
{\cal T}^\La =\left (\begin{array}{cc} T^\La & 0\\ 0 &
-(T^\La)^*\end{array}\right ),
\eeq{16}
where $T^\La$ is a hermitian generator of the gauge group. This equation
implies that, for example, if $x$ transforms in the fundamental representation
of the gauge group, $y$ transforms in the anti-fundamental. The Killing
vectors are:
\beq
k^u_\La = -{i\over 2} \tr\bar e^u {\cal T}_\La Q.
\eeq{17}
The three complex structures are
\beq
(J^x)^u_v = {-i\over 2} \tr\bar e^u\sigma^xe_v.
\eeq{18}
They correspond to the left multiplication of the quaternion by $-i\vec\sigma$
and therefore satisfy the quaternionic algebra.

In the flat limit, the covariant $SU(2)$ derivative simplifies to the ordinary
differential in the equation for the prepotential:
\beq
2k^u_\La h_{uw}\vec J^w_v = \partial_v\vec P_\La ,
\eeq{19}
which can be solved to give
\beq
\vec P_\La = {1\over 2}\tr\vec\sigma Q^\dagger{\cal T}_\La Q .
\eeq{20}
Notice that for an Abelian gauge field we have:
\beq
{\cal T} = q\sigma_3.
\eeq{21}
It will be sometimes useful to define the vector
\beq
{\cal Q} = \left (\begin{array}{c} x\\ y\end{array}\right ).
\eeq{22}

As for the gauge-field Lagrangian, we  choose for the physical fields
the sections
\beq
(X^a,F_a)=\left({1\over \sqrt{2}}g\phi^a,
{-i\over\sqrt{2}g}g_{ab}\phi^b\right),
\eeq{220}
where $g_{ab}=\tr (T_aT_b)$
and define $\phi = \phi^aT_a$.
Expanding the K\"ahler potential,
the bosonic part of the Lagrangian (for a single factor in the gauge group and
only one hypermultiplet in the fundamental) reduces to
\beq
{\cal L} = -{1\over g^2}\tr F_{\mu\nu}F^{\mu\nu} + \tr \nabla\phi
\nabla\phi^\dagger +{1\over 2}\tr\nabla{\cal Q}^\dagger \nabla{\cal Q} -
V(\phi,
{\cal Q}),
\eeq{22'}
with the potential:
\beq
V(\phi,
{\cal Q}) = g^2\left (\tr\left (\left [\phi,\phi^\dagger\right ]\right )^2
 + \tr{\cal Q}^\dagger\left \{\phi, \phi^\dagger\right\} {\cal Q}
+ \vec P_a\vec P_b(g^{-1})_{ab}\right ).
\eeq{23}
Any dependence on the frozen moduli $z^i$ has been re-absorbed in the
normalization for $\phi$.
One can recognize the standard rigid N=2 gauge Lagrangian. The triplet of
prepotentials has reduced to the triplets of D-terms of N=2 supersymmetry.

\section{Soft Breaking Terms and Mirror Splitting}
A concrete example of softly broken N=2 (rigid) supersymmetry with tree-level
mirror fermion mass splitting was given in~\cite{DA}; in this section, we
review that example. The model in~\cite{DA} has a gauge group $SU(3)\times
SU(4)\times U(1)$, and the physical quarks and leptons are arranged in
representations of the gauge group as in eq.~(\ref{2}). The Higgs sector
responsible for the breaking of $SU(4)\times U(1)$ is made of a scalar field,
$\phi$, partner of the $SU(4)\times  U(1)$ gauge field
under N=2 supersymmetry, and four complex scalars $(x_i, y_i)$, $i=1,2$,
arranged in two hypermultiplets. The potential simplifies if we include the
coupling constants dependence in $\Phi$
\beq
\Phi_i=g_4 T^I_i\phi_I + g_1 Z_i\phi,
\eeq{DA00}
where $T^I_i$ are the generators of $SU(4)$ and $Z_i$ is the generator of
$U(1)$. This definition will be used only in the minimization of the
potential. Obviously, in the kinetic term the fields $T^I_i\phi_I,\;\phi$
appear
separately.
The normalizations are as in formula~(\ref{22'}). The index $i=1,2$ in the
definition above labels
the different representations of the gauge group acted upon by $\Phi$; we
shall omit it wherever unnecessary.
The first hypermultiplet transforms in the
$(1,4,+1/2)\oplus (1,\bar{4},-1/2)$ of the gauge group, while the second
transforms in the $(1,4,-1/2)\oplus (1,\bar{4},+1/2)$.
The leptons and quarks (physical and mirror) get their tree-level masses only
from the term
\beq
\overline{X}_L\Phi Y_L, \;\;\; \overline{X}_Q\Phi Y_Q.
\eeq{DA0}
In order to give a large mass to the mirrors, while keeping the physical
fermions light, the VEV of $\Phi$ must be off-diagonal. This never happens with
a pure N=2 potential.

Indeed, the
N=2 supersymmetric scalar potential depends on these fields as follows:
\bea
V_{N=2}(x_i,y_i,\Phi) &=& \Big\{{1\over g_4^2}\tr([\Phi,\Phi^\dagger])^2
+\sum_i
(x^\dagger\{\Phi_i^\dagger,\Phi_i\}x_i +
y_i^\dagger\{\Phi_i^\dagger,\Phi_i\}y_i)\Big\} +
\nonumber \\
&& g_4^2\Big\{{1\over 4}\sum_{ij}|x^\dagger_i x_j +y^\dagger_i y_j|^2 +{1\over
2}\sum_{ij}
\{(y^\dagger_jy_i)(x^\dagger_jx_i) -(y^t_jx_i)(x^\dagger_iy^*_j)\} +
\nonumber \\
&& -{1\over 16} \big(\sum_i |x_i|^2 +\sum_i |y_i|^2\big)^2 +{1\over 4}
\big(\sum_i|x_i|^2\big)\big(\sum_j|x_j|^2\big) + \nonumber \\
&& -\big(\sum_i y^t_ix_i\big)\big(\sum_j x^\dagger_jy^*_j\big)\Big\}
+{1\over 8}\Big\{\big[\sum_i q_i(|x_i|^2-|y_i|^2)\big]^2 + \nonumber \\
&& 4\sum_{ij}q_iq_j (y^t_ix_i)(x^\dagger_jy_j^*)\Big\}.
\eea{DA1}
Here $q_1=1$, $q_2=-1$, and normalizations are as in formula~(\ref{22'}) for
each factor in the gauge group.

In the absence of soft breaking terms, the potential is always non-negative,
and it has a global minimum ($V_{N=2}=0$) at
\beq
x_i=y_i=0, \;\;\; [\Phi^\dagger,\Phi]=0.
\eeq{DA2}
The vanishing of the commutator implies that $\Phi$ is diagonal up to a gauge
rotation.

An off-diagonal VEV for $\Phi$ can be obtained by introducing appropriate
soft terms (scalar masses and scalar tri-linear couplings) which explicitly
break the rigid N=2 supersymmetry, but preserve the most important property
of supersymmetry, namely, the absence of quadratic divergences~\cite{GG}.
These terms are:
\beq
V_{soft}=\sum_i\Big[m_{x_i}^2|x_i|^2 + m_{y_i}^2|y_i|^2 +
m_{\Phi_i}^2|\Phi_i|^2 +\Big(
M_iy^t_i\Phi_i x_i+ c.c.\Big)\Big].
\eeq{DA3}
The mass parameters $m_{x_i}$, $m_{y_i}$,$m_{\Phi_i}$,
and $M_i$ (which are complex, in general) can be adjusted to give VEVs of
arbitrary magnitude to $x_i$, $y_i$, and $\Phi$. The minimization of the
potential $V_{N=2}+V_{soft}$ is
arduous, for arbitrary values of $V_i\equiv |\langle x_i \rangle|$,
$v_i\equiv |\langle y_i \rangle|$, and $g_4\hat{v}\equiv |\langle \Phi
\rangle|$,
but it becomes doable in the
approximation $V_i \gg \hat{v} \gg v_i$~\cite{DA}.

In this limit, there is only one term $O(V_i^4)$ in the potential which
contributes to the alignment of $x_i$: $1/2|x^\dagger_1x_2|^2$.
It favors $x_i \perp x_2$, so that we can choose
\beq
\langle x_1\rangle =
\left(\begin{array}{c} 0\\ 0\\ V_1 \\ 0 \end{array}\right), \;\;\;
\langle x_2\rangle =
\left(\begin{array}{c} 0\\ 0\\ 0\\ V_2 \end{array}\right).
\eeq{DA4}
The first VEV breaks $SU(4)\times U(1)$ to $SU(2)_L\times SU(2)_R \times U(1)$
while the second breaks this group to $SU(2)\times U(1)$.

The next largest terms which contribute to alignment are $O(V_i^2\hat{v}^2)$,
and read:
\beq
\sum_i \Big[ x^\dagger_i\{\Phi_i^\dagger,\Phi_i\}x_i + \Big(
M_iy^t_i\Phi_i x_i +c.c.\Big)\Big].
\eeq{DA5}
By consistency, $M_i=O(g_4V_i\hat{v}/v_i)\gg V_i$.
The minimization of eq.~(\ref{DA5}) with respect to $\Phi$ can be done exactly
since $\Phi$ appears there quadratically, and gives
\bea
&&\;\Phi_{a3}=-{M_1^*\over V^*} y^*_{1\, a},\;\;\;\;\;
\Phi_{a4}=-{M_2^* \over V^*}
y^*_{2\, a},
\nonumber\\
&& \;\Phi_{43}=-{M^*_1\over 2V^*} y^*_{1\, 4}, \;\;\;\;\;
\Phi_{34}=-{M^*_2 \over 2V^*} y^*_{2\, 3}, \nonumber \\
&& \Phi_{1\, 33}=-{M^*_1\over 2V^*}y^*_{1\, 3} ,\;\;\; \Phi_{2\, 44}=
-{M^*_2\over 2V^*} y^*_{2\, 4},\nonumber \\
&& \;\Phi_{3a}=\Phi_{4a}=0 ,\;\;\;\;\;\;\;\; a,b=1,2.
\eea{DA6}
Here, for simplicity, and without loss of generality, we assumed
$V_1=V_2\equiv V$.
The potential term in eq.~(\ref{DA5}), computed at the stationary point for
$\Phi$ given in eq.~(\ref{DA6}), gives rise to a {\em negative} mass term for
$y_i$:
\beq
- \sum_i\left({1\over 2}| M_i y_{i\, 3}|^2 +{1\over 2}|M_i y_{i\, 4}|^2\right)
 - \sum_{ai}M_i^2|y_{i\, a}|^2.
\eeq{DA7}
For an appropriate {\em positive} value of the mass term $m^2_{y_i}$ in
$V_{soft}$, the minimum in $y_i$ becomes
\beq
y_{i\, 3}=y_{i\, 4}=0, \;\;\; y_{i\, a}\neq 0.
\eeq{DA8}

Finally, the only $O(\hat{v}^4)$ alignment term in the potential,
$\tr [\Phi^\dagger\Phi]^2$, implies that, in the presence of small, positive
mass terms $m_{\Phi_i}^2$, the potential $V_{N=2}+V_{soft}$ is
minimized by an off-diagonal VEV of $\Phi$, which, by an $SU(2)_L$
rotation, can be brought into the form
\beq
\langle\Phi\rangle=g_4\left( \begin{array}{cccc} 0 & 0 & \hat{v}_1 & 0 \\
                                0 & 0 & 0 & \hat{v}_2 \\
                                0 & 0 & 0 & 0 \\
                                0 & 0 & 0 & 0 \end{array} \right).
\eeq{DA9}
The off-diagonal elements of $\Phi$ play here the role of the standard-model
Higgs, and their VEVs $\hat{v}_i$ break $SU(2)\times U(1)$ to $U(1)$. The
electric charge is $Q=T_{3L}+T_{3R}+Z$, where $T_{3L}$, $T_{3R}$ are the
diagonal generators of $SU(2)_L$, $SU(2)_R$, and $Z$ is $1/2(B-L)$ on matter
fields.

The $\Phi$ in eq.~(\ref{DA9}) generates a mass $M_{U}=\sqrt{2}\hat{v}_1g_4$
for the mirror up quarks and leptons, while the mirror down quarks and leptons
have a mass $M_D\sqrt{2}\hat{v}_2g_4$. The constraint on the mirror masses
eq.~(\ref{1}) now becomes, more
precisely:
\beq
M_U^2+ M_D^2 =2M_W^2.
\eeq{DA10}
Additional sub-dominant terms in the potential ($O(v^2\hat{v}^2)$ etc.) may
generate
small nonzero VEVs for the block-diagonal components of $\Phi$. These terms
induce small mixing between mirror and physical fermions. These mixings are
not in contradiction (indeed, they are favored) by experimental data~\cite{NR}.
\section{The Supersymmetry Breaking Mechanism}
The simplest method to break rigid supersymmetry is to introduce
Fayet-Iliopoulos
(FI) terms for a $U(1)$ gauge fields in the theory. In an N=2 gauge theory one
can introduce three FI terms
corresponding to constant pieces in the triplet of D-terms. The D-terms
appear in the supersymmetric transformation
formula for the gaugino in a combination which, roughly speaking, is
$\delta\lambda = \vec D\vec\sigma\epsilon$. A nonzero expectation value for
$\vec D$ breaks supersymmetry.

In the local case the D-terms are replaced by the prepotentials $\vec P^\La$,
which appear in the gaugino and gravitino shifts. For
example, the gravitino shift reads:
\beq
\delta \psi_{A\, \mu}=-{1\over 2}e^{K/2} (\vec\sigma)_A^{\; C} \epsilon_{BC}
\vec
P_\Lambda X^\Lambda(z)\gamma_{\mu}\eta^B\equiv
iS_{AB}\gamma_{\mu}\eta^B.
\eeq{A}
A nonzero value for $\vec
P_\Lambda X^\Lambda(z)$ breaks supersymmetry. We choose
prepotentials and sections in such a way  that $\vec
P_\Lambda X^\Lambda(z)$ has an (essentially) constant complex piece.
Let us remind the reader that we need to break supersymmetry with two
different arbitrary scale (two different masses for the two gravitinos), in
order to break the global $SU(2)$ R-symmetry. This is required because we
want to give different masses for the scalars $x_i$ and $y_i$ of Section 2,
and they are doublets under the R-symmetry.
The gravitino mass matrix is $S_{AB}$; by a suitable choice of the
complex number
$\vec
P_\Lambda X^\Lambda(z)$ it can have two arbitrary eigenvalues.

A minimal supergravity model with all these characteristics was constructed
in~\cite{FGP}, where it was used to partially break N=2 supersymmetry
to N=1 in Minkowsky space.
In this model, the charged-hypermultiplet scalars parametrise the
quaternionic manifold $SO(4,1)/SO(4)$,
and the vector-multiplet scalars parametrise the K\"ahler
manifold $SU(1,1)/U(1)$.

Let us denote the quaternionic coordinates of the hypermultiplet manifold
by $b^u$, $u=0,1,2,3$.
The metric is  $h_{uv}=(1/2{b^0}^2)\delta_{uv}$ and the three complex
structures are exactly as in eq.~(\ref{19}).
The quaternionic potentials read:
\beq
\omega^x_u={1\over b^0}\delta^x_u, \;\;\; \Omega^x_{0u}=-{1\over
2{b^0}^2}\delta^x_u, \;\;\; \Omega^x_{yz}={1\over 2
{b^0}^2}\epsilon^{xyz},\;\; x,y,z=1,2,3.
\eeq{f}
The manifold is invariant under
arbitrary constant translation of the coordinates $b^1,b^2,b^3$. We can then
choose constant Killing vectors $k_\La^u = \delta_{ux}\zeta^x_\La$,
where $\zeta^x_\La$ are arbitrary constants.

The prepotentials can be determined using the formulas of Section 1, and
they read:
\beq
P^x_\La = {1\over b_0}\zeta^x_\La .
\eeq{g}
The $\vec{P}_\Lambda$ are real functions, nonzero and independent of the vector
multiplets. Excluding the dependence on $b_0$, which becomes irrelevant in
the flat limit, they can be considered as three real constants,
exactly as FI terms are expected to be.

An undesired dependence on $z$ in $\vec
P_\Lambda X^\Lambda(z)$ can be avoided using the sections
\beq
X^0(z) = -{1\over \sqrt{2}},\;\;\; X^1(z) = {i\over \sqrt{2}},\;\;\;
F_0 ={i\over \sqrt{2}} z,\;\;\;  F_1 = {1\over \sqrt{2}}z .
\eeq{d}
This choice gives the $SU(1,1)/U(1)$ K\"ahler potential
\beq
K = -\log (z +\bar z).
\eeq{e}
Note that no prepotential exists for such a choice of sections~\footnotemark.
\footnotetext{One can find these sections by the symplectic
transformation (electric-magnetic duality)
$X^1\rightarrow -F_1$, $F_1\rightarrow X^1$ of the basis
specified by the prepotential
$F(X^\Lambda)=iX^0X^1$.}
The absence of prepotential is necessary in order to get a partial
breaking to N=1 at zero cosmological constant.
We will make the same choice of sections, since it simplifies the formulas and
avoids a $z$ dependence in the sections.

With an appropriate choice of the complex vector
\beq
{\cal \vec M}=X^0\vec P_0+X^1\vec
P_1={1\over\sqrt{2}}(\vec\zeta_0 +i\vec\zeta_1),
\eeq{444}
the gravitino mass matrix,
$\vec\sigma{\cal \vec M}\sigma_2$, can be given two arbitrary eigenvalues.
In this way we can break supersymmetry to N=0 with two arbitrary scales.
\section{The Supergravity Model}
In this section we construct a supergravity model which, in the flat
limit, reproduces the soft breaking terms discussed in Section 2 and
lifts the mass degeneracy between mirror and physical fermions.
We work in
the approximation in which supergravity formulas reduce to those of rigid
supersymmetry in the physical sector, and the kinetic term of the physical
fields is canonical.
In other
words, we keep only the first term of the supergravity Lagrangian in the
$1/M_P$ expansion.

Let us start the description of our model.

Let us begin with the quaternionic manifold of the hypermultiplets.
In the full supergravity model at the scale $M_P$ the
hidden and physical hypermultiplets parametrise a complicated
quaternionic manifold, which in the flat limit reduces to the
product of the hidden-sector quaternionic manifold ($SO(4,1)/SO(4)$)
times quaternions with flat metric.
The question of whether this quaternionic manifold exists is answered in
Appendix A, where such space is explicity constructed.

The multiplets of our model are those of the rigid N=2 theory
described in Section 2, coupled through gravitational
interaction to the hidden sector
described in Section 3. The only thing we need to remember about the hidden
sector is that the quaternionic manifold $SO(4,1)/SO(4)$ admits three
independent ``translations'' as
isometries, which can be used to break supersymmetry by giving a (complex)
constant term to the quantity $\vec
P_\Lambda X^\Lambda(z)$.
To these fields, we add a
hypermultiplet $(\tilde x, \tilde y)$ with a negative kinetic term, neutral
under the
physical gauge group and a non-propagating, auxiliary, Abelian vector
multiplet.
The two Higgs hypermultiplets $(x_i, y_i)$ have charges $q_i$
under the $U(1)$ vector of this multiplet while
$(\tilde x, \tilde y)$ has charge $p$, and the scalars in $SO(4,1)/SO(4)$
translate by the constant vector $\vec h$. The introduction of non-propagating
gauge fields in N=2 supergravity is also known as quaternionic
quotient~\cite{Gal} and it is one of the most powerful ways to construct new
quaternionic
manifolds from known ones. Since this auxiliary gauge field has no kinetic
term, it can be eliminated, together with its supersymmetric partners,
using their equation of motion. In particular, $\vec{P}_{aux}=0$, allows
to express
$(\tilde x, \tilde y)$ in terms of $(x_i, y_i)$ up to a $U(1)$ transformation.
The action of the non-dynamical $U(1)$ is manifestly free on the manifold
defined by the equation $\vec{P}_{aux}=0$, thus, the quotient manifold is
regular~\cite{Gal}. Despite the negative kinetic term for $(\tilde x, \tilde
y)$, crucial for reasons which will be soon explained, the quotient manifold
has a positive definite metric.

Next, we choose the sections:
\bea
X^0 &=& -{1\over\sqrt{2}},\;\;X^1 = {i\over\sqrt{2}},\;\; X^{aux} =
{1\over\sqrt{2}}\Phi^{aux},\;\;X^a = g{1\over\sqrt{2}}\Phi^a,\nonumber\\
F_0 &=& {i\over\sqrt{2}}z,\;\;F_1 = {1\over\sqrt{2}}z,\;\; F_{aux} =
0,\;\;F_a = -{i\over\sqrt{2}g}g_{ab}\Phi^b.
\eea{50}
where $g_{ab}=\tr (T_aT_b)$. The indices $0,1$ label the graviphoton and the
hidden $U(1)$, while the
index $a$ labels the physical gauge group~\footnotemark.
\footnotetext{In formulas~(\ref{50},\ref{51}) only one factor in the gauge
group is indicated. Our theory has, obviously, three coupling constants,
$g_1$,$g_4$, and the $SU(3)$ gauge coupling.}
The K\"ahler potential reads (after reintroducing the powers of $M_P$ for
the fields in the physical sector):
\beq
K = -\log\left ( z +\bar z -
{\Phi^a\bar\Phi^b\tr (T_aT_b)\over M_P^2}\right ).
\eeq{51}
Expanding in power of $M_P$, the first nontrivial term in $\Phi$ gives the
standard normalization for the gauge fields and their partners.

We have (using eqs.~(\ref{20}) and~(\ref{g}))
\beq
X^\La \vec P_\La = {{\cal\vec M}\over M_P b_0} + \sum_i
{1\over 2\sqrt{2}M_P^3}\tr\vec
\sigma Q^\dagger_i \left (\begin{array}{cc} \Phi_i & \;\\ \; &
-\Phi^T_i\end{array}\right )Q_i.
\eeq{52}
Notice that the constant contribution from the hidden sector has been
chosen of order
$1/M_P$, to insure non-vanishing
interference with the physical sector. The auxiliary gauge
field and  $(\tilde x, \tilde y)$ do not appear in eq.~(\ref{52}),
since $\vec{P}_{aux}=0$, but they
will appear in the potential through the terms involving $k_{aux}$.
 From now on, we will no longer indicate the powers of $M_P$. The flat limit
corresponds to
$|\vec{h}|\gg Q,\Phi$.

Now, expand the potential in eq.~(\ref{12}).
We obtain the remarkably simple formula
\bea
V = && g_4^2\tr\left (\left
[\Phi,\Phi^\dagger\right ] \right )^2 + g_4^2\vec P^{(4)}_a\vec P^{(4)}_b
(g^{-1})_{ab} + 4g_1^2\vec P^{(1)}\vec P^{(1)}\nonumber\\
&&-
2\left |X^\La \vec P_\La\right |^2 +
4 h_{uv} k^u_\La k^v_\Sigma X^\La X^\Sigma .
\eea{53}
Any dependence on $z+\bar z$ has been re-absorbed in a rescaling of $\Phi$ and
${\cal M}$.
The first two terms in~(\ref{53}) were already present in eq.~(\ref{23}),
and are in the standard
rigid Lagrangian for gauge fields and hypermultiplets. The fourth term contains
the missing term in eq.~(\ref{23}), needed to complete the rigid Lagrangian.
The rest of the third and fourth terms is the interference
between the hidden and the physical sector and, thus, gives
the soft breaking terms we are looking for.

Let us collect the expressions for the Killing vectors (cfr.
eqs.~(\ref{17},\ref{21}))
\bea
SO(4,1)/SO(4)&:&\;\;\;\;\; X^\La
k^u_\La = \delta^{ux}\left ({\cal M}^x + {\Phi^{aux}\over\sqrt{2}}h^x
\right ) ,\nonumber\\
\tilde Q &:& \;\;\;\;\; X^\La k^u_\La e_u=
-ip{\Phi^{aux}\over\sqrt{2}}\sigma_3\tilde Q ,\nonumber\\
Q_i &:& \;\;\;\;\; X^\La k^u_\La e_u = -iq_i{\Phi^{aux}\over\sqrt{2}}\sigma_3
Q_i
-i{1\over\sqrt{2}}\left ( \begin{array}{cc} \Phi & \;
\\ \; & -\Phi^T\end{array} \right)Q_i .
\eea{54}
We see that the square of the second term in the Killing vector for $Q$ exactly
reproduces the missing term in the rigid Lagrangian.

The interference terms in the scalar potential are:
\bea
&-& {1\over \sqrt{2}}\sum_i
\tr{{\cal\vec M}^*\over b_0}\vec\sigma Q_i^\dagger \left (\begin{array}{cc}
\Phi & \;\\ \; & -\Phi^T\end{array}\right )Q_i +
\Phi^{aux}\left [{\sqrt{2}\over b_0^2}{\cal\vec M}^*
\vec h + 2\sum_iq_i\left (x_i^\dagger\Phi^\dagger x_i + y^t_i\Phi^\dagger
y^*_i\right ) + h.c. \right]
\nonumber\\
&+& |\Phi^{aux}|^2\left [{\vec h^2\over b_0^2} +
\sum_i2q_i^2\left ( |x_i|^2 + |y_i|^2\right ) -
2p^2\left ( |\tilde x|^2 + |\tilde y|^2\right ) \right ].
\eea{56}
Our aim is to reproduce exactly all the soft breaking terms in
formula~(\ref{DA3}).
The first term in~(\ref{56}) can reproduce
the tri-linear coupling in formula~(\ref{DA3}), by correctly orienting the
vector ${\cal\vec M}$. We will choose
\beq
{{\cal\vec M}^*\over b_0}\vec\sigma = \left (\begin{array}{cc} \epsilon & B\\ 0
&
-\epsilon\end{array}\right )
\eeq{55}
If $\epsilon =0$ we get exactly and only the desired coupling in~(\ref{DA3}).
So we can identify B with $M=O(V\hat v/v)$, the largest scale in our theory.
The need for $\epsilon$ will be clear soon.

The other terms in formula~(\ref{DA3}) are generated by the quaternionic
quotient.
As explained before, the auxiliary field $\Phi^{aux}$ in formula~(\ref{56})
can be eliminated using its equation of motion and the result must be
evaluated on the submanifold:

\beq
\vec{P}^{aux} =
{\vec h\over b_0} + \sum_i{q_i\over 2}\tr\vec\sigma Q_i^\dagger\sigma_3 Q_i
- {p\over 2}\tr\vec\sigma \tilde Q^\dagger\sigma_3 \tilde Q = 0.
\eeq{57}

This produce soft breaking terms, among which the mass terms for $x_i$ and
$y_i$, with coefficents determined by $p$ and $q_i$.

The generation of a positive mass term for $\Phi$ is more subtle. In the
derivation of formula~(\ref{53}) we assumed that the spontaneous symmetry
breaking
occurs with zero cosmological constant. This is true for the model discussed in
section 3, but it is no longer true after taking
the quaternionic quotient, which
generates a cosmological constant $M_P^2E_0$. By expanding the factor $e^K$ in
front of the potential in formula~(\ref{12}), we see that we generate the
mass term $E_0\Phi^2$. The sign of $E_0$, as we will see, is determined by
the sign of the kinetic term for $\tilde x$ and $\tilde y$. We choose a
negative metric for $\tilde x$ and $\tilde y$ just to generate a positive
mass term for $\Phi$.

Let us derive the explicit expression for the soft breaking terms.
By keeping only the relevant terms in the $M_P$ expansion in
formula~(\ref{57}), and orienting
$\vec h$ in the direction 3 ($\vec h = (0,0,h)$), we find
\beq
|\tilde x|^2 + |\tilde y|^2 = \sqrt{{1\over 4}\left (\tr\vec\sigma\tilde
Q^\dagger\sigma_3\tilde Q\right )\left (\tr\vec\sigma\tilde
Q^\dagger\sigma_3\tilde Q\right )} \approx {h\over pb_0} + \sum_i{q_i\over
p}\left
( |x_i|^2 - |y_i|^2 \right ),
\eeq{58}
where we have chosen the same sign for $h$ and $p$.

Substituting this expression back in eq.~(\ref{56}), eliminating the
auxiliary
field $\Phi^{aux}$, and expanding the resulting expression, we finally get
\bea
-\sqrt{2}By^t\Phi x &-& \sqrt{2}\epsilon^*\sum_i\left [\left ({2q_ihb_0\over
h^2-2phb_0}
+ 1
\right )\left (x_i^\dagger\Phi x_i\right )
+ \left (-{2q_ihb_0\over h^2-2phb_0}
+ 1
\right )\left (
  y_i^t\Phi y_i^*\right )\right ] + h.c.
\nonumber \\ &+&
\sum_i\left
|{2q_ihb_0\epsilon \over h^2-2phb_0}\right |^2\left [ 2\left ( 1 - {p\over
q_i}\right
)|x_i|^2  +
2\left ( 1 + {p\over q_i}\right )|y_i|^2\right ] + E_0\tr\Phi^2.
\eea{59}
The cosmological constant reads
\beq
E_0 = - {2|\epsilon |^2h^2\over (h^2-2phb_0)(z+\bar z)}
\eeq{cc}
We can use the three free parameters $\epsilon , p,q_i$ to extablish the
hierarchy of scales $V\gg \hat v\gg v$.
The extra parameter $h/b_0$ can be fixed in order to get a positive
cosmological
constant and, as a consequence, a positive mass for $\Phi$. A convenient
limit is $p\gg h/b_0$, which gives a positive cosmological constant,
whose magnitude can be made as small as desired.

We get finally,
\bea
\left[ -\sqrt{2}By\Phi x - \sqrt{2}\epsilon^*\sum_i \left(
1 -{q_i\over p}\right )\left(x_i^\dagger\Phi x_i\right )
- \sqrt{2}\epsilon^*\sum_i \left(
1 +{q_i\over p}\right )\left(
  y_i^t\Phi y_i^*\right)
+ h.c. \right] +
\nonumber\\+2|\epsilon |^2\sum_i
\left |{q_i\over p}\right |\left [
\left ( 1 - {q_i\over p}\right )|x_i|^2  +
\left ( 1 + {q_i\over p}\right )|y_i|^2\right ] + {\epsilon^2h\over
b_0p(z+\bar z)}\tr\Phi^2.
\eea{60}

As discussed in Section 2,
we need a large positive mass term for $y_i$, of order
$O(g_4 V\hat v/v)$, the same as the tri-linear coefficient
$y^t_i\Phi x_i$, and a
negative mass for $x_i$, of order $O(g_4 V)$. This can be easily achieved by
choosing
$B,\epsilon=O(g_4 V\hat v/v)$, and by tuning
${q_i/p} = 1 + O(v^2/\hat v^2)$. The term $x_i^\dagger\Phi x_i$ is
an undesired one; including it in the minimization of Section 2,
would completely change the alignments. Fortunately, the condition that
this term is suppressed with respect to the good tri-linear term,
\beq
\left|By^t_i\Phi x_i\right| \gg
\left|\epsilon^*\left (1 - {q_i\over p}\right )x^\dagger_i\Phi x_i\right|,
\eeq{61}
gives the condition $\hat v^2\gg vV$. This condition was not originally
present in~\cite{DA}, but can be easily satisfy without interfering with
phenomenological constraints on $\hat{v}$ and $V$, namely,
$\hat{v}=O(100\, GeV)$, $V\gg 1\, TeV$.
\section{Conclusions and Comments}
In this paper, we have presented an N=2 supergravity model where supersymmetry
is spontaneous broken in a hidden sector at a scale $\sqrt{MM_P}$ ,
$M\equiv V\hat{v}/v$. The
hidden sector communicates only through interactions of gravitational strength
with an observable sector. In the flat limit, supersymmetry breaking affects
the observable sector through the appearance of soft
terms that explicitly break the {\em rigid} N=2 supersymmetry, and trigger the
breakdown of both the standard-model $SU(2)\times U(1)$, and the symmetry
between physical fermions and mirrors. This mechanism is the N=2 analog of the
hidden-sector supersymmetry breaking in N=1 supergravity~\cite{N}.

Our soft supersymmetry breaking terms are those of ref.~\cite{DA}
plus some small sub-dominant terms. The most important of them are the
tri-linears
\beq
\left(1 - {q_i\over p}\right)\epsilon^* x_i\Phi_i x_i.
\eeq{con1}
The presence of such terms induces small nonzero VEVs for $\Phi_{ab}$,
$a,b=3,4$, of order  $|(1-q/p)\epsilon|=Vv/\hat{v}\ll \hat{v}$.
These nonzero VEVs are not dangerous; indeed, they may even be beneficial to
the model since they induce small mixing angles between physical fermions and
mirrors. As recalled in Section 2, these mixing angles are favored by recent
experimental data~\cite{NR}.

The details of our construction involve two ingredients: the choice of a
particular realization of the ``special K\"ahler geometry'' for
the N=2 vector
multiplets, and the extensive use of the technique of quaternionic quotients to
define an appropriate manifold for the hypermultiplets.

As explained in Sections 3, 4, we define the special geometry of the
vector multiplets by giving $n_V+1$ holomorphic sections such that no
prepotential exists. This choice evades an old no-go
theorem~\cite{CGP2} which forbids spontaneous breaking of N=2 supergravity to
N=1 in flat space. This same choice of sections proves useful here, even
though in the present case we do not have any argument to show that it is
necessary.

The use of quaternionic quotients is a powerful method to find a
``custom made'' quaternionic space for the hypermultiplets. This technique is
particularly useful in the case where the quaternionic manifolds are
non-compact. This is the physically interesting one and, luckily, in this
case the technique does not run into the snags that mar its application to
the construction of compact manifolds (see~\cite{Gal} for details).

A comment about the cosmological constant and the radiative stability of this
model is in order. The
introduction of an auxiliary Abelian vector multiplet induces a tree-level
cosmological constant of order $M_P^2m_\Phi^2$ ($m_\Phi$ is defined
in Section 2).
This is not a serious problem, since in any case
radiative corrections to the cosmological constant are
$O(M_P^2M^2)\gg M_P^2m_\Phi^2$: N=2 neither helps solving nor worsens
the cosmological constant problem. A more serious problem
is that the hierarchy of scales introduced in our model may be destabilized by
radiative corrections. This is an important problem well worth investigating;
the fact that our effective action is defined at the scale $V\hat{v}/v \ll
M_{GUT}$ may render this problem less severe.

Finally, let us comment on the uniqueness of our model. We have not proven
that our is the unique way of constructing an N=2 supergravity without light
fermions; indeed, the message is the opposite: N=2 supergravity in its most
general formulation is more flexible a theory than generally supposed, and it
can easily account for a realistic particle spectrum. More general models may
conceivably be constructed, which have a zero tree-level cosmological
constant, or that, more importantly, give rise to an N=2 grand unified model.
\newpage
\noindent
{\bf Acknowledgements}
\vskip .1in
\noindent
L.G. would like to thank the Department of Physics of NYU for
its kind hospitality. L.G. is supported in part by the
Ministero dell' Universit\`a e della Ricerca Scientifica e Tecnologica,
by INFN and by the European Commission TMR program
ERBFMRX-CT96-045, in which L.G. is associated to the University of
Torino.
M.P. is supported in part by NSF under grant no. PHY-9318171.
A.Z. is supported in part by DOE grant no. DE-FG02-90ER40542 and by the
Monell Foundation.
\section*{Appendix A: The Quaternionic Manifold}
\renewcommand{\theequation}{A.\arabic{equation}}
\setcounter{equation}{0}
In our paper we assumed that a quaternionic
manifold with some special properties exists. In this appendix, we explicitly
construct that manifold, using the method of quaternionic quotients introduced
in ref.~\cite{Gal}.

As we saw in the paper, to construct our N=2 model we need a
quaternionic manifold which reduces to $SO(4,1)/SO(4)$ when the scalars of
all Higgs, quark, and lepton hypermultiplets are set to zero. On the other
hand, the manifold cannot be an $SO(5,n)/SO(4)\times SO(1,n)$~\footnotemark,
since this coset structure implies a doubling of quark and lepton generations.
\footnotetext{As explained in
the text, our construction requires a manifold with indefinite metric.
The metric is positive definite on the subspace defined by equation~(\ref{57}).
} This result comes
about since in $SO(5,n)/SO(4)\times SO(1,n)$ the hypermultiplets belong to
{\em real} representations of the gauge group. Indeed, in this case, the coset
representatives can be written as $Q^M_\mu$, where the index $\mu$ labels the
fundamental of $SO(4)$ and $M$ labels the vectorial of $SO(1,n)$, and the
allowed representations of the physical gauge group must be contained in the
vectorial of $SO(1,n)$.
For instance, each lepton hypermultiplet
(physical plus mirror) in the $(1,4,+1/2)\oplus (1,\bar{4},-1/2)$ is
paired with another hypermultiplet in the $(1,4,-1/2)\oplus (1,\bar{4},+1/2)$.
We want instead a manifold where some coordinates are represented as in
eq.~(\ref{15'}) without further constraints.

A quaternionic manifold with the desired properties can be constructed as a
quaternionic quotient of the space
\beq
{\cal M} \subset USp(2,n)/USp(1)\times USp(1,n).
\eeq{G1}
This space can be represented by $n+2$ quaternionic homogeneous coordinates
$Q_0,Q_1,...,Q_{n+1}$, identified modulo left multiplication by unit
quaternions
($Q_i\sim UQ_i$, $U^\dagger U=1$, $i=0,..,n+1$),
and subject to the constraints
\bea
-Q_0^\dagger Q_0 + \sum_{i=1}^{n} Q^\dagger_i Q_i -Q^\dagger_{n+1} Q_{n+1} &=&
-1. \nonumber \\
-Q_0^\dagger Q_0 + \sum_{i=1}^{n} Q^\dagger_i Q_i &<& 0.
\eea{G2}
Our quaternionic quotient is defined by coupling the fields $Q_0,..,Q_k$,
$k<n+1$
to a non-dynamical $SU(2)$ gauge field acting by right multiplication with
unit quaternions:
\beq
Q_i\rightarrow Q_iV,\;\;\; i=0,..,k,\;\;\; V^\dagger V=1.
\eeq{G3}
The resulting quaternionic space is $M\equiv M_0/SU(2)$, where $M_0$ is the
algebraic submanifold of ${\cal M}$ defined by setting
to zero the prepotentials of the non-dynamical $SU(2)$~\cite{Gal}:
\beq
M_0=\{ Q\in {\cal M}|
-Q_0\sigma_IQ_0^\dagger  +  \sum_{i=1}^k Q_i\sigma_IQ_i^\dagger=0\}.
\eeq{G4}
A subtlety arises at this point: the action of SU(2) on $M_0$ is not free, and
the quotient space $M$ has orbifold singularities.
This is most easily seen by defining the non-homogeneous coordinates
\beq
Q_i= Q_0q_i, \;\;\; q_0=1,\;\;\; Q_0^\dagger Q_0=
1/(1-\sum_{i=1}^{n} q^\dagger_iq_i + q^\dagger_{n+1}q_{n+1}).
\eeq{G5}
The $SU(2)$ action induced on the $q_i$s by eq.~(\ref{G3}) is
\beq
q_i\rightarrow V^{-1}q_iV, \;\;\; i=0,..,k,\;\;\; q_i\rightarrow V^{-1}q_i,
\;\;\; i=k+1,..,n+1.
\eeq{G6}
On the subspace of $M_0$ where $q_i$ vanishes for all $i>k$, the center of
$SU(2)$, $Z_2$,
acts trivially, while when $q_i\neq 0$ for some $i>k$, the
group acts freely. Moreover, it can be shown that the isotropy group of
$M_0$ is always either $Z_2$ or the identity~\cite{Gal}.
This result implies that our space $M$ is a $Z_2$ orbifold,
with a singularity at $q_i=0$, $i>k$. It is easy to check this statement
explicitly using our coordinates $q_i$.
In terms of them, the constraint in eq.~(\ref{G5})
becomes independent of $q_i$, $i>k$:
\beq
-\sigma_I + \sum_{i=1}^k q_i\sigma_Iq_i^\dagger=0.
\eeq{G7}
The quotient space $M_0/SU(2)$ can be described very explicitly by using
the $SU(2)$
gauge invariance to transform one of quaternions $q_i$, $i=1,..,k$, into
the diagonal form $a+ib\sigma_3$. This ``gauge fixing'' leaves only the
center $Z_2$ as residual symmetry, since by the constraint eq.~(\ref{G7}) not
all $q_i$ are diagonal.
We must still divide the resulting space by $Z_2$.
This means that topologically $M$ is an open subset of
\beq
\big[SO(4,k-3)/SO(4)\times SO(k-3)\big]\times H^{(1,n-k)}/Z_2,
\eeq{G8}
where $Z_2$ acts on  $H^{(1,n-k)}$ (the quaternionic hyperplane of dimension
$n+1-k$ and signature $(1, n-k)$) as:
\beq
q_i\rightarrow -q_i   ,\;\;\; i>k.
\eeq{G9}
Moreover, since the metric of $M$ near $q_i=0$, $\forall i>k$,
factorises into
the metric of $SO(4,k-3)/SO(4)\times SO(k-3)$ times the {\em flat} metric of
$H^{(1,n-k)}$, the singularity at the origin of $H^{(1,n-k)}$ is an orbifold.

Unlike the case of compact spaces studied in~\cite{Gal}, here it
is trivial to find a smooth manifold associated to $M$; it is
sufficient to remove the identification given in eq.~(\ref{G9})! The
resulting space, $\tilde{M}$ is the $Z_2$ covering of $M$ ($M=\tilde{M}/Z_2$),
and is manifestly smooth and quaternionic. Obviously, $M$ is not metrically
complete, i.e. it is a part of a larger, possibly singular,
quaternionic manifold. This is not a
problem for us, since only a small neighborhood of the (smooth) point $q_i=0$,
$i=1,..,n$, is relevant to the flat limit used in our construction.

\end{document}